# Influence analyses of "designs" for evaluating inconsistency in network meta-analysis

Kotaro Sasaki, MPH

The Graduate Institute for Advanced Studies, The Graduate University for Advanced Studies (SOKENDAI), Tokyo, Japan

Human Biology Integration Foundation, Deep Human Biology Learning, Eisai Co. Ltd., Tokyo, Japan

ORCID: https://orcid.org/0009-0007-3081-9438

Hisashi Noma, PhD[*]

Department of Interdisciplinary Statistical Mathematics, The Institute of Statistical Mathematics, Tokyo, Japan

The Graduate Institute for Advanced Studies, The Graduate University for Advanced Studies (SOKENDAI), Tokyo, Japan

ORCID: http://orcid.org/0000-0002-2520-9949

*Corresponding author: Hisashi Noma

Department of Interdisciplinary Statistical Mathematics

The Institute of Statistical Mathematics

10-3 Midori-cho, Tachikawa, Tokyo 190-8562, Japan

TEL: +81-50-5533-8440

e-mail: noma@ism.ac.jp

**Abstract**

Network meta-analysis is an evidence synthesis method for comparing the effectiveness of multiple available treatments. To justify evidence synthesis, consistency is an important assumption; however, existing methods founded on statistical testing can be substantially limited in statistical power or have several drawbacks when handling multi-arm studies. Moreover, inconsistency can be theoretically explained as design-by-treatment interactions, and the primary purpose of such analyses is to prioritize the further investigation of specific "designs" to explore sources of bias and other issues that might influence the overall results. In this article, we propose an alternative framework for evaluating inconsistency using influence diagnostics methods, which enable the influence of individual designs on the overall results to be quantitatively evaluated. We provide four new methods, the averaged studentized residual, MDFFITS, $\Phi_d$, and $\Xi_d$, to quantify the influence of individual designs through a "leave-one-design-out" analysis framework. We also propose a simple summary measure, the $O$-value, for prioritizing designs and interpreting these influential analyses in a straightforward manner. Furthermore, we propose another testing approach based on the leave-one-design-out analysis framework. By applying the new methods to a network meta-analysis of antihypertensive drugs and performing simulation studies, we demonstrate that the new methods accurately located potential sources of inconsistency. The proposed methods provide new insights into alternatives to existing test-based methods, especially the quantification of the influence of individual designs on the overall network meta-analysis results.

## 1. Introduction

Network meta-analysis is an evidence synthesis method for analyzing the comparative effectiveness of multiple available treatments by pooling evidence from direct and indirect comparisons [1,2]. To justify evidence synthesis in network meta-analysis, consistency is an important assumption. Conventionally, "consistency" refers to the agreement between the evidence of direct and indirect comparisons [1]; however, Higgins et al. [3] showed that this concept can be rigorously explained as design-by-treatment interactions in the network, where in this case, "design" refers to the combination of treatments compared in the corresponding studies. In addition, "inconsistency" refers to a disagreement in the treatment effects across different combinations of treatment comparisons in the network, and various statistical tools to evaluate it have been developed [3-7].

Currently, most inconsistency evaluation methods are founded on statistical testing and have substantial limitations in accuracy or other drawbacks. Representative testing approaches include local and global inconsistency tests [3,4]. Local testing evaluates the concordance of effect sizes on a specified triangular loop of treatments. The main drawback of local methods is that there is no unique way to handle multi-arm studies ($\geq 3$ arms), which can lead to different estimates and conclusions [8]. Moreover, their statistical power to detect loop inconsistency is limited because a small number of studies are involved in individual triangles in practice [9,10]. By contrast, a global inconsistency test evaluates the inconsistency on an entire network (i.e., it tests the design-by-treatment interactions in the network) [3]. This approach enables multi-arm studies to be included in a valid way and uses all statistical information on a network to detect inconsistency. However, this approach also has limited statistical power in practice because most individual designs involve only a few studies (i.e., the interaction test is founded on the partially "sparse" nature of the dataset) [11]. Furthermore, even if inconsistency is detected, this testing method does not provide precise information about where the differences in the effect exist. Other test-based inconsistency evaluation methods have similar drawbacks, especially statistical-power limitations due to the sparsity of the datasets in practice. However, the primary purpose of an inconsistency evaluation is not to provide deterministic conclusions. Rather, its purpose is to identify possible sources of bias or irregular problems in the network from the perspective of discordances in effect sizes among the different combinations of compared treatments. After checking the results of inconsistency tests, we usually conduct further investigations to determine clinical or methodological factors that could be influencing the design-by-treatment interactions. Given these issues, current testing approaches might not be suitable for such analyses; instead, alternative methodological frameworks that enable the prioritization of further investigations and uncertainty evaluations might be more useful in practice.

In this article, we propose new methods to prioritize study designs that indicate possible design-by-treatment interactions for further investigations. To do this, we adopt an alternative framework for assessing the inconsistency of network meta-analysis. In particular, we adopt the influence diagnostics framework and provide new methods to assess the influence of a design on the basis of how it changes the overall results of the network meta-analysis via leave-one-design-out analysis. We also provide methods to quantify the uncertainty of the influence measures; these methods enable the certainty of the change in the overall results to be quantitatively and statistically evaluated. The proposed methods are more suitable than other methods for prioritizing individual designs and quantifying their influence, which makes them suitable as preliminary analyses for further investigations. In addition, these methods can handle multi-arm studies in a simple manner and directly identify the sources of inconsistency in practice. We illustrate the effectiveness of these new methods by applying them to evaluate a network meta-analysis of antihypertensive drugs [12] and further evaluate their performance in simulation studies.

## 2. Models and inference methods for network meta-analysis

We first describe the contrast-based approach for network meta-analysis, which models relative treatment effect measures as multivariate outcomes [3,11]. Suppose that $N$ studies are involved in the analysis and that $p + 1$ treatments are compared. Let $Y_{ij}$ denote an estimate of the treatment effect for treatment $j$ compared with a reference treatment (e.g., a placebo) in study $i$ ($i = 1,2, \dots, N$; $j = 1,2, \dots, p$) using, for example, the log odds ratio or mean difference. Moreover, let $\boldsymbol{Y_i} = (Y_{i1}, Y_{i2}, \dots, Y_{ip})^T$ denote the multivariate outcome. The multivariate statistical model for contrast-based network meta-analysis is given as

$$\boldsymbol{Y_i} \sim \text{MVN}(\boldsymbol{\theta_i}, \boldsymbol{S_i}), \tag{1}$$

where $\boldsymbol{S_i}$ (a $p \times p$ matrix) denotes the within-study covariance matrix, defined as

$$\boldsymbol{S_i} = \begin{pmatrix} s_{i1}^2 & \rho_{i12}s_{i1}s_{i2} & \cdots & \rho_{i1p}s_{i1}s_{ip} \\ \rho_{i21}s_{i2}s_{i1} & s_{i2}^2 & \cdots & \rho_{i2p}s_{i2}s_{ip} \\ \vdots & \vdots & \ddots & \vdots \\ \rho_{ip1}s_{ip}s_{i1} & \rho_{ip2}s_{ip}s_{i2} & \cdots & s_{ip}^2 \end{pmatrix},$$

which is usually assumed to be known and fixed to a valid estimate [13]. In addition, $s_{ij}$ is the standard error corresponding to $Y_{ij}$, and $\rho_{ijk}$ denotes the pairwise correlation coefficient between the outcomes for treatments $j$ and $k$. Furthermore, $\boldsymbol{\theta_i}$ denotes the true underlying effects for the $i$th study; that is, $\boldsymbol{\theta_i} = (\theta_{i1}, \theta_{i2}, \dots, \theta_{ip})^T$. When a study does not include a reference treatment, the data augmentation approach can be used, in which pseudo-data containing minimal information are added into the reference arm (e.g., 0.001 events for 0.01 patients) [11]. Note that in the data augmentation approach, the choice of reference category does not affect the estimates of the treatment effects.

Furthermore, we adopt the random-effects model to address between-studies heterogeneity, which is expressed as

$$\boldsymbol{\theta}_i \sim \text{MVN}(\boldsymbol{\mu}, \boldsymbol{\Sigma}),$$

where $\boldsymbol{\mu} = (\mu_1, \mu_2, \dots, \mu_p)^T$ denotes the average treatment effect parameters and $\boldsymbol{\Sigma}$ is the between-studies covariance matrix, expressed as

$$\boldsymbol{\Sigma} = \begin{pmatrix} \tau_1^2 & \kappa_{12}\tau_1\tau_2 & \cdots & \kappa_{1p}\tau_1\tau_p \\ \kappa_{21}\tau_2\tau_1 & \tau_2^2 & \cdots & \kappa_{2p}\tau_2\tau_p \\ \vdots & \vdots & \ddots & \vdots \\ \kappa_{p1}\tau_p\tau_1 & \kappa_{p2}\tau_p\tau_2 & \cdots & \tau_p^2 \end{pmatrix}.$$

Because there are rarely sufficient studies available for all of the variance–covariance parameters in $\boldsymbol{\Sigma}$ to be identified, the equal variance assumption is adopted in most cases ($\tau^2 = \tau_1^2 = \tau_2^2 = \cdots = \tau_p^2$) [6,14]. Under the equal variance assumption, all pairwise correlation coefficients $\kappa_{jk}$s should be equal to 0.5 because consistency is assumed [15]. We also adopt this assumption. In addition, let $\boldsymbol{W}_i = (\boldsymbol{S}_i + \boldsymbol{\Sigma})^{-1}$ denote the inverse of the marginal covariance matrix of $\boldsymbol{Y}_i$. When $\tau^2 = 0$, this model reduces to the fixed-effect model [16].

To estimate the parameters, restricted maximum likelihood (REML) estimation is widely adopted. The REML estimator $(\hat{\boldsymbol{\mu}}, \hat{\boldsymbol{\Sigma}})$ is obtained by maximizing the log-restricted likelihood function

$$\ell(\boldsymbol{\mu}, \boldsymbol{\Sigma}) \propto \sum_{i=1}^{N} \left\{ \log\{\det(\boldsymbol{W}_i^{-1})\} + (\boldsymbol{Y}_i - \boldsymbol{\mu})^T \boldsymbol{W}_i (\boldsymbol{Y}_i - \boldsymbol{\mu}) \right\} + \log\left\{ \det\left( \sum_{i=1}^{N} \boldsymbol{W}_i \right) \right\}.$$

The covariance matrix estimator of $\hat{\boldsymbol{\mu}}$ is given as $\hat{V}[\hat{\boldsymbol{\mu}}] = \left( \sum_{i=1}^{N} \widehat{\boldsymbol{W}}_i \right)^{-1}$, where $\widehat{\boldsymbol{W}}_i = (\boldsymbol{S}_i + \hat{\boldsymbol{\Sigma}})^{-1}$. Because most individual clinical trials involve only two to four arms, some elements of $\boldsymbol{Y}_i$ are often undefined and regarded as missing. In such cases, we replace $\boldsymbol{Y}_i$ and $\boldsymbol{S}_i$ with their subvectors and submatrices in the estimating functions [13]. The Wald-type tests and confidence intervals for $\boldsymbol{\mu}$ are constructed using the asymptotic normality of the REML estimator [16]. Several other effective inference methods involving higher-order asymptotic approximation are available for this multivariate random-effects model [14,16], and the following discussion can be adapted to address these alternative inference methods.

### 3. Influence diagnostics of designs

When assessing the inconsistency of a network meta-analysis, we consider prioritizing the study designs that indicate possible design-by-treatment interactions for further investigation. The key idea is that researchers should carefully consider the designs that have a particularly strong influence on the overall results in the primary analysis. Even if the inconsistency tests are significant, if the corresponding designs only weakly influence the overall results, they will not be important factors when interpreting the evidence of the network meta-analysis. Therefore, we propose adopting the influence diagnostics framework to quantitatively assess, via leave-one-design-out analysis, the influence of individual study designs on the basis of how they change the overall results of a network meta-analysis. Influence diagnostic methods for individual studies have been well established [17-20]; however, our concern is assessing the influence of a subset of studies with the same design. In this section, we provide four effective methods for evaluating the influence diagnostics of designs: the averaged studentized residual, MDFFITS, $\Phi_d$, and $\Xi_d$. These proposed methods evaluate influences from different perspectives: the residual, difference in fit, and relative changes in measures of heterogeneity. They are founded on the general influence diagnostics methodology established for various statistical models [17-20]. Furthermore, the selected methods have been well studied over a long time, and their general characteristics are comprehensively reviewed in existing papers and textbooks [21-23].

*3.1 Averaged studentized residual*

We first propose an inconsistency diagnostic measure based on conventional residual-based statistics [21]. The studentized residual standardized by the standard error is one of the most commonly used measures to assess the influence of individual subjects [17,19-21]. However, in this case, we consider assessing the influence of a set of studies with the same design. Because this assessment involves multiple studies, the conventional studentized residual cannot be directly adopted. To quantify the influence of multiple studies simultaneously, we propose using the "averaged" studentized residual.

Let $D$ denote the total number of designs in the network and $T_d$ denote a set of studies with design $d$ $(d = 1, \ldots, D)$. We further use $(\widehat{\boldsymbol{\mu}}^{(-T_d)}, \widehat{\boldsymbol{\Sigma}}^{(-T_d)})$ to denote the REML estimator for the leave-one-design-out dataset that excludes the studies with the $d$th design, and $\widehat{\boldsymbol{W}}_i^{(-T_d)} = (\boldsymbol{S}_i + \widehat{\boldsymbol{\Sigma}}^{(-T_d)})^{-1}$. We then first define the multivariate version of a studentized residual for the $i$th study as a quadratic form scaled by the number of comparisons $p_i$ $(i = 1, \ldots, N)$ as follows:

$$\xi_i = \frac{1}{p_i}(\boldsymbol{Y}_i - \widehat{\boldsymbol{\mu}}^{(-T_d)})^T \widehat{V}[\boldsymbol{Y}_i - \widehat{\boldsymbol{\mu}}^{(-T_d)}]^{-1}(\boldsymbol{Y}_i - \widehat{\boldsymbol{\mu}}^{(-T_d)}),$$

where $\widehat{V}[\boldsymbol{Y}_i - \widehat{\boldsymbol{\mu}}^{(-T_d)}] = (\widehat{\boldsymbol{W}}_i^{(-T_d)})^{-1} + (\sum_{k \in (-T_d)} \widehat{\boldsymbol{W}}_k^{(-T_d)})^{-1}$. Note that, in the definition of $\xi_i$, the reference treatment should be set to one of the treatment arms in the $i$th study to avoid computational irregularities (the reference should be switched if the reference treatment was not included in that study). We hence define the averaged studentized residual for the $d$th design as the mean of $\xi_i$ of the studies with design $d$ as follows:

$$\Psi_d = \frac{1}{N_d} \sum_{i \in T_d} \xi_i,$$

where $N_d$ refers to the number of studies of design $d$. This influence measure is interpreted as a summary measure that assesses the overall influence of the studies with design $d$ and can be simply used as the conventional studentized residual. In general, a large value of $\Psi_d$ indicates that a large difference exists between the estimates from studies with design $d$ and the overall estimates from the other studies; that is, design $d$ could be a source of inconsistency and has a strong influence that could change the overall results, and hence it should be carefully considered for further investigation. This influence measure can also be used as a ranking measure to prioritize designs for inconsistency assessment.

Another relevant problem is quantifying the statistical error of $\Psi_d$ to assess the size of the realized values and provide a reasonable threshold for deciding whether further investigation should be considered. To solve this issue, we propose a parametric bootstrap approach. The proposed algorithm for estimating the sampling distribution of $\Psi_d$ is presented in Algorithm 1.

*Algorithm 1: Bootstrap for estimating the sampling distribution of $\Psi_d$*

Step 1. Under the multivariate random effects model (1), which assumes global consistency (no design-by-treatment interactions) in the network, compute the REML estimates of $(\widehat{\boldsymbol{\mu}}, \widehat{\boldsymbol{\Sigma}})$ using the all-studies dataset.

Step 2. Resample $\boldsymbol{Y}_1^{(b)}, \boldsymbol{Y}_2^{(b)}, \dots, \boldsymbol{Y}_n^{(b)}$ from the estimated distribution of (1) using $(\widehat{\boldsymbol{\mu}}, \widehat{\boldsymbol{\Sigma}})$ for the parameters via parametric bootstrap, $B$ times $(b = 1,2,\dots,B)$.

Step 3. Compute the average studentized residual $\Psi_d^{(b)}$ for the $b$th bootstrap sample $\boldsymbol{Y}_1^{(b)}, \boldsymbol{Y}_2^{(b)}, \dots, \boldsymbol{Y}_n^{(b)}$. Next, replicate this process for all $B$ bootstrap samples and calculate $\Psi_d^{(1)}, \Psi_d^{(2)}, \dots, \Psi_d^{(B)}$.

We can then estimate the sampling distribution of $\Psi_d$ from the bootstrap distribution of $\Psi_d^{(1)}, \Psi_d^{(2)}, \dots, \Psi_d^{(B)}$ under the consistency assumption. Comparing the realized value of $\Psi_d$ and the bootstrap distribution, we can quantitatively evaluate how the actual dataset diverges from the overall results when the consistency assumption is fulfilled. We consider the corresponding design to be influential if the realized value of $\Psi_d$ exceeds a certain quantile of the bootstrap distribution (e.g., the upper 5% quantile) [18,21]. In this work, we propose a convenient measure, the *O*-value, which can be used to quantify the extremeness of the realized value in the bootstrap distribution intuitively and is formulated as follows:

$$O = \frac{1}{B} \sum_{b=1}^{B} I\left(\Psi_d \leq \Psi_d^{(b)}\right),$$

where $I(\cdot)$ is an indicator function. A small *O*-value indicates that the realized value is extreme under the consistency assumption and that the corresponding design could be influential. The threshold can be set arbitrarily (e.g., 5%), and we can screen individual designs in the same sense with conventional regression diagnostics using this summary measure [21]. Because a certain number of designs should be evaluated in a network meta-analysis, the *O*-value is a useful measure for interpreting the overall results intuitively and immediately. Furthermore, to prioritize the designs on the basis of the influence measure, the *O*-values can be used as a ranking measure. The *O*-values can be similarly defined for the other influence measures proposed in the following sections.

### 3.2 Difference-in-fit measure

Another statistical measure widely used in conventional influence diagnostics is the difference-in-fit (DFFITS) measure [17,21]. The DFFITS quantifies the discrepancy between the overall estimate based on all subjects and that based on a leave-one-out analysis, which is defined by the change in an estimate, standardized by a variation measure. For the inconsistency evaluation task, we consider the influence of a set of studies with the same design; that is, we propose a DFFITS measure defined by a leave-one-design-out analysis.

For the multivariate random-effect model (1), we focus on the overall estimate of $\boldsymbol{\mu}$ and propose a multivariate DFFITS (MDFFITS) as follows:

$$\text{MDFFITS}_d = \frac{1}{p_d}\left(\widehat{\boldsymbol{\mu}} - \widehat{\boldsymbol{\mu}}^{(-T_d)}\right)^T \widehat{V}\left[\widehat{\boldsymbol{\mu}}^{(-T_d)}\right]^{-1}\left(\widehat{\boldsymbol{\mu}} - \widehat{\boldsymbol{\mu}}^{(-T_d)}\right),$$

where $p_d$ is the number of treatment comparisons for design $d$. Note that the reference treatment should also be set to one of the treatment arms in design $d$ to avoid computational irregularities and should be switched if it is not. MDFFITS quantifies how the overall estimate of $\boldsymbol{\mu}$ is changed by the leave-one-design-out analysis, which is standardized by the variance estimate of $\widehat{\boldsymbol{\mu}}^{(-T_d)}$ and the number of treatment comparisons. If MDFFITS increases, the corresponding design $d$ strongly influences the overall estimates, suggesting a design-by-treatment interaction. To quantify the statistical error, we can also adopt the parametric bootstrap approach given in Algorithm 2.

*Algorithm 2: Bootstrap for estimating the sampling distribution of* $\text{MDFFITS}_d$

Step 1. Perform Steps 1 and 2 of Algorithm 1 to generate bootstrap samples $\boldsymbol{Y}_1^{(b)}, \boldsymbol{Y}_2^{(b)}, \dots, \boldsymbol{Y}_n^{(b)}$ $(b = 1, 2, \dots, B)$.

Step 2. Compute the DFFITS statistic $\text{MDFFITS}_d^{(b)}$ for the $b$th bootstrap sample $\boldsymbol{Y}_1^{(b)}, \boldsymbol{Y}_2^{(b)}, \dots, \boldsymbol{Y}_n^{(b)}$. Next, replicate this process for all $B$ bootstrap samples and calculate $\text{MDFFITS}_d^{(1)}, \dots, \text{MDFFITS}_d^{(B)}$.

We can then obtain the bootstrap distribution of $\text{MDFFITS}_d$ and determine the *O*-value comparing the realized value with the bootstrap distribution. A ranking based on the *O*-values can be used to prioritize the study designs, and the realized values can be a useful reference for considering further investigation.

*3.3 Relative changes in heterogeneity statistics*

Between-studies heterogeneity reflects the discordance in the effect sizes of individual studies. If a strong design-by-treatment interaction exists in the network, one set of studies will have effect sizes that are substantially different from those of other sets, and the heterogeneity statistics will indicate a large between-studies heterogeneity in general. In this sense, the concept of heterogeneity is related to the concept of inconsistency [7]. In particular, if there is a strong design-by-treatment interaction that warrants particular attention, the heterogeneity statistics would change markedly. Thus, the relative change in the heterogeneity statistics caused by leave-one-design-out analysis will be an effective measure for assessing the design-by-treatment interaction. A similar approach has been adopted in outlier analyses for various meta-analysis methods [17,18,20].

First, we consider the variance $\tau^2$ in the covariance matrix of the random-effects distribution, which represents the heterogeneity. A measure of the relative change in the $\tau^2$ estimate is defined as the ratio of the variance in the leave-one-design-out analysis to the variance of the data of all studies, as follows:

$$\Phi_d = \frac{\hat{\tau}^{2\,(-T_d)}}{\hat{\tau}^2},$$

where $\hat{\tau}^{2\,(-T_d)}$ is the REML estimator of $\tau^2$ for the leave-one-design-out analysis. Parameter $\Phi_d$ quantifies the impact of design $d$ on the size of the estimate of the between-studies variance $\tau^2$. When $\Phi_d$ is less than 1, the heterogeneity in the network decreases after design $d$ has been excluded, suggesting that design $d$ could contribute to the between-studies heterogeneity in the network (i.e., design $d$ could have a design-by-treatment interaction). In particular, a large change in $\Phi_d$ can indicate a strong design-by-treatment interaction. Conversely, when $\Phi_d$ is greater than 1, the heterogeneity increases after exclusion. In general, this scenario rarely happens and is not usually related to design-by-treatment interactions.

Second, we consider the $I^2$ statistic. The $I^2$ statistic was originally proposed by Higgins and Thompson [24] for conventional pairwise meta-analysis and was generalized to the multivariate meta-analysis by Jackson et al. [25]. This measure has also been widely used in numerous systematic reviews as an effective tool for assessing heterogeneity in network meta-analysis [2]. The $I^2$ statistic for multivariate meta-analysis [6] is calculated as

$$I^2 = \max\left(0, \frac{R^2 - 1}{R^2}\right), R = \det\left(\frac{V[\hat{\boldsymbol{\mu}}_R]}{V[\hat{\boldsymbol{\mu}}_F]}\right)^{\frac{1}{2p}},$$

where $\hat{\boldsymbol{\mu}}_F$ and $\hat{\boldsymbol{\mu}}_R$ are the common effects and grand mean estimators for the fixed-effects and random-effects models, respectively; the fixed-effect model corresponds to model (1) constrained by $\tau^2 = 0$. We propose an alternative influence measure that is defined as the change in the $I^2$ statistic in the leave-one-design-out analyses relative to that of the all-studies data, calculated as follows:

$$\Xi_d = \frac{I^{2(-T_d)}}{I^2},$$

where $I^{2(-T_d)}$ is the $I^2$ statistic for the leave-one-design-out analysis. The interpretation is the same as that of $\Phi_d$. When $\Xi_d$ is less than 1, the heterogeneity in the network decreases after design $d$ has been excluded, suggesting that design $d$ could have a design-by-treatment interaction; in particular, a large change in $\Xi_d$ could indicate a strong design-by-treatment interaction. Note that Jackson et al. [25] provided another definition of the $I^2$ statistic using the $H^2$ statistic, and the influence measure $\Xi_d$ can be defined similarly using this alternative definition of $I^2$.

To quantify the statistical error, we can apply the parametric bootstrap approach, as given in Algorithm 3.

*Algorithm 3: Bootstrap for estimating the sampling distribution of $\Phi_d$ and $\Xi_d$*

Step 1. Perform Steps 1 and 2 of Algorithm 1 to generate bootstrap samples $\boldsymbol{Y}_1^{(b)}, \boldsymbol{Y}_2^{(b)}, \dots, \boldsymbol{Y}_n^{(b)}$ $(b = 1, 2, \dots, B)$.

Step 2. Compute the $\Phi_d$ and $\Xi_d$ statistics, $\Phi_d^{(b)}$ and $\Xi_d^{(b)}$, for the $b$th bootstrap sample $\boldsymbol{Y}_1^{(b)}, \boldsymbol{Y}_2^{(b)}, \dots, \boldsymbol{Y}_n^{(b)}$. Next, replicate this process for all $B$ bootstrap samples and calculate $\Phi_d^{(1)}, \dots, \Phi_d^{(B)}$ and $\Xi_d^{(1)}, \dots, \Xi_d^{(B)}$.

In this way, we can the obtain the bootstrap distributions of $\Phi_d$ and $\Xi_d$ and determine the *O*-values comparing the realized values with the bootstrap distributions. We can also create ranked lists based on the *O*-values to prioritize the study designs, and the *O*-values can be used as a reference for further investigation.

## 4. Alternative testing-based approach

In Section 3, we presented influence diagnostics-based approaches for the evaluation of design-by-treatment interactions based on leave-one-design-out analysis. In this section, a similar, straightforward approach based on statistical testing using the leave-one-design-out analysis is considered. Krahn et al. [7] discussed a simple Wald statistic under a fixed-effect model that assesses the concordance of the common effects parameters of a set of studies with design $d$ with those of other studies. Their method can be extended to the random-effects model (1). Let $\boldsymbol{\mu}^{(T_d)}$ and $\boldsymbol{\mu}^{(-T_d)}$ denote the grand mean vectors of the two subsets. We can assess their possible discordance using a statistical test. The problem is formulated as

$$\text{H}_0: \boldsymbol{\mu}^{(T_d)} - \boldsymbol{\mu}^{(-T_d)} = \boldsymbol{0} \text{ v.s. } \text{H}_1: \boldsymbol{\mu}^{(T_d)} - \boldsymbol{\mu}^{(-T_d)} \neq \boldsymbol{0}.$$

The Wald statistic for assessing the null hypothesis $H_0$ is given as

$$W_d = \left(\hat{\boldsymbol{\mu}}^{(Td)} - \hat{\boldsymbol{\mu}}^{(-Td)}\right)^T \hat{V}^{-1} \left(\hat{\boldsymbol{\mu}}^{(Td)} - \hat{\boldsymbol{\mu}}^{(-Td)}\right),$$

where the covariance matrix estimate $\hat{V} = \hat{V}\left[\hat{\boldsymbol{\mu}}^{(Td)} - \hat{\boldsymbol{\mu}}^{(-Td)}\right] = \sum_{i \in (T_d)} \widehat{W}_i^{(Td)} + \sum_{i \in (-T_d)} \widehat{W}_i^{(-Td)}$. Wald statistic $W_d$ asymptotically follows a $\chi^2$-distribution with $p_d$ degrees of freedom under the null hypothesis. When the null hypothesis is rejected, this suggests a discordance in effect sizes (i.e., the inconsistency of design $d$). Notably, the results of the Wald test accord with those of the inconsistency tests based on side-splitting [5,26,27] if the corresponding design $d$ is a two-arm design. Although side-splitting focuses on specific treatment pairs, our proposed testing approach focuses on specific designs.

From a mathematical perspective, the large sample approximation using a $\chi^2$-distribution might be violated in a case with a limited number of studies [28]. Even in such cases, the parametric bootstrap approach discussed in Section 3 can be used to estimate the sampling distribution of $W_d$, and the resultant inference can be improved [28]. The bootstrap algorithm is given in Algorithm 4.

*Algorithm 4: Bootstrap for estimating the sampling distribution of $W_d$*

Step 1. Perform Steps 1 and 2 of Algorithm 1 to generate bootstrap samples $Y_1^{(b)}, Y_2^{(b)}, \ldots, Y_n^{(b)}$ $(b = 1,2, \ldots, B)$.

Step 2. Compute the Wald statistic $W_d^{(b)}$ for the $b$ th bootstrap sample $Y_1^{(b)}, Y_2^{(b)}, \ldots, Y_n^{(b)}$. Next, replicate this process for all $B$ bootstrap samples and calculate $W_d^{(1)}, \ldots, W_d^{(B)}$.

We can determine the bootstrap $P$-value using the empirical distribution of $W_d^{(1)}, \ldots, W_d^{(B)}$ as the reference distribution instead of the $\chi^2$-distribution. Notably, this testing approach suffers from the limited statistical power and instability of $\tau^2$ estimation in subset $T_d$ when only a few studies are considered; however, in practice, it might be a useful option familiar to many practitioners.

## 5. Application

To illustrate the effectiveness of the proposed methods, we applied them to a network meta-analysis of antihypertensive drugs by Sciarretta et al. [12]. The network meta-analysis included 26 randomized controlled trials comparing a placebo and seven drug classes: α-blocker (AB), angiotensin-converting enzyme inhibitor (ACE), angiotensin II receptor blocker (ARB), β-blocker (BB), calcium channel blocker (CCB), conventional treatment (CT), and diuretic (DD). The outcome was the incidence of heart failure. The network diagram of this study is presented in Figure 1, and the 26 studies and their designs are summarized in Table 1. The network included 18 designs, two of which were three-arm trials (STOP-2 and ALLHAT2002); the others were all two-arm trials. Note that the Jikei Heart Study included in the design of ARB vs CT was found to include falsified data, and the main result paper was retracted [29]; the active treatment was categorized as an ARB, and its efficacy was reported to be much larger than that of CT. Thus, in this network, the efficacy of ARB could be overestimated because of this extreme data. Given this fact, a naïve hypothesis is that an ARB vs CT design could raise a design-by-treatment interaction in this network.

The primary results of this network meta-analysis using the REML method are shown in Figure 2(a). The odds ratio (OR) was adopted as the measure of effect, and the Placebo group was set as the reference. For the between-study heterogeneity, $\hat{\tau}$ was 0.099 and $I^2$ was 56.9%. In addition, the $P$-value of the global inconsistency test was 0.459; significant inconsistency was not detected, possibly because of the low statistical power.

We used the four influence diagnostics methods for this network meta-analysis. Note the AB vs DD design was excluded in the analyses because AB was included in only one study (ALLHAT2000) and was isolated in the network. In the bootstrap analyses, we performed 5,000 resamplings consistently and formally regarded the designs with $O <$ 0.05 as influential designs. The results are presented in Figure 3. The averaged studentized residuals and the $O$-values for the 17 designs are presented in Figure 3(a) and (b); the design IDs are presented in Table 1. DD vs Placebo (design 17), ARB vs CT (design 11), and ACE vs DD (design 7) had the largest averaged studentized residuals, and the $O$-values of the first two designs were less than 0.05. ARB vs CT included the Jikei Heart Study and E-COST, and DD vs Placebo included HYVET. MDFFITS gave results similar to those of the averaged studentized residuals for the $O$-values, as shown in Figure 3(c) and (d). Although the rankings of MDFFITS were slightly different, the $O$-values of the ARB vs CT and DD vs Placebo designs were less than 0.05. In addition, the results obtained by the analyses using $\Phi_d$ and $\Xi_d$ are presented in Figure 3(e)–(h). Both $\Phi_d$ and $\Xi_d$ were less than 1 for seven designs (DD vs Placebo, ARB vs Placebo, ARB vs CT, BB vs CCB, CCB vs Placebo, ACE vs DD, and ARB vs BB). Unlike the former two methods, only DD vs Placebo showed $O < 0.05$. Moreover, the overall rankings of both measures based on heterogeneity statistics were similar.

Table 2 shows the results of the leave-one-design-out analysis tests. The $P$-values were computed using the bootstrap method with 5,000 resamplings, and the level of significance was set to 0.05. The statistical tests detected only ARB vs CT as a significant design (the bootstrap $P$-value was less than 0.05). The pooled OR estimates and 95% confidence intervals (CIs) were 1.501 (0.991, 2.272) for studies with the ARB vs CT design and 0.911 (0.744, 1.117) for the other studies.

In summary, the DD vs Placebo design was detected by all four influence diagnostics methods. This design included the HYVET study, which is a trial studying the effectiveness of DD for patients 80 years of age or older. The background of the patients in this study is substantially different from those of the other studies. Furthermore, this study was terminated at the interim analysis because of the efficacy of DD [30], and hence the effect of DD might have been overestimated (OR: 0.375, 95% CI: 0.228, 0.615). In addition, ARB vs CT was detected by the two influence diagnostics methods using averaged studentized residual, MDFFITS, and the proposed testing approach. This design included the Jikei Heart Study and E-COST trial. As expected, the Jikei Heart Study had an extreme profile in the network, and the proposed methods detected the corresponding design as an influential one that could change the overall results.

To assess the potential influence of the two detected study designs on the overall estimates, we conducted a sensitivity analysis that excluded the corresponding three studies: Jikei Heart Study, E-COST, and HYVET. The pooled results are presented in Figure 2(b). After the studies were excluded, all OR estimates increased except for that of CT. In particular, the efficacy of ARB decreased and the rankings of ARB and CT were reversed. These results imply that the treatment effects of all the investigated drugs except for CT might be overestimated in the original network meta-analysis. Moreover, the heterogeneity decreased: the $\tau$ estimate changed from 0.099 to 0.054, and $I^2$ changed from 56.9% to 31.7%. Furthermore, the $P$-value of the global inconsistency test increased from 0.459 to 0.911. In the original report of this network meta-analysis [12], the inconsistency was reported to be low; however, these results provide additional insights, and the overall results should be interpreted more carefully.

## 6. Simulations

We conducted simulation studies to evaluate the performances of the proposed methods under practical network meta-analysis scenarios. Simulation data were generated to mimic the antihypertensive drugs dataset used in Section 5. First, we generated the incidence of events for placebo arm $p_{i0}$ for the $i$th study from a uniform distribution Uni$(0.006, 0.167)$, and the log ORs between the other treatments and placebo $\boldsymbol{\theta_i}$ were generated from a multivariate normal distribution MVN$(\boldsymbol{\mu}, \boldsymbol{\Sigma})$ $(i = 1,2, \dots, N)$. In addition, $\boldsymbol{\mu}$ was set to the log OR estimates obtained from the real-data analysis in Section 5 that excluded the three trials that corresponded to possibly inconsistent designs; that is, $\boldsymbol{\mu} = (0.264, -0.294, -0.210, -0.072, -0.129, -0.276, -0.441)^T$. Covariance matrix $\boldsymbol{\Sigma}$ was defined as a variance-covariance matrix with diagonal elements equal to $\tau^2$ and off-diagonal elements equal to $\tau^2/2$ ($\tau = 0.05$ or $0.10$). Inconsistency was introduced by adding an inconsistency factor $\omega$ to the corresponding component of $\boldsymbol{\theta_i}$ as $\theta_{ij}^{inc} = \theta_{ij} + \omega$ for the selected designs $(j = 1,2, \dots, p)$. Factor $\omega$ was set to $0, -0.1$, or $-0.3$ to represent no inconsistency, moderate inconsistency, and severe inconsistency, respectively. Two designs (ARB vs CT and CCB vs CT) were selected to be inconsistent designs; the latter corresponded to a three-arm trial. For individual scenarios, one of the two designs was selected, and the inconsistency factor was added to the log ORs of the ARB or CCB arms of the corresponding design. Then, the incidence of events for the $j$th treatment arm in the $i$th study was calculated as $p_{ij} = p_{i0} \exp(\theta_{ij})/\{p_{i0}\exp(\theta_{ij}) + 1 - p_{i0}\}$; for studies where inconsistency was introduced, $\theta_{ij}^{inc}$ was used instead of $\theta_{ij}$. The sample size $n_{ij}$ for the $i$th study was assumed to be equal among all the arms of the individual studies and was randomly drawn from integers between 204 and 15,268. Furthermore, the number of studies $N$ was set to 26 or 52. When $N$ was 26, the study designs followed the structure of the real data in Section 5 (presented in Table 1); when $N$ was 52, the number of studies per design was doubled. Finally, the number of events in the $j$th arm $d_{ij}$ was generated from a binomial distribution Bin$(n_{ij}, \, p_{ij})$.

Next, we calculated the contrast-based summary statistics from the simulated datasets and analyzed the resultant data using the proposed methods (the four influence diagnostics methods and the testing approach). For comparison, we adopted the local inconsistency test [4], which is a widely used method for assessing inconsistency. We replicated the simulations 1,000 times. For the 1,000 simulations, we calculated the true positive rate (TPR) for the scenarios with inconsistency and the false positive rate (FPR) for the scenarios without inconsistency; the FPR conceptually corresponds to the type I error rate in the hypothesis testing framework. In addition, we assessed the rate at which the $P$-values or $O$-values were below 5% for the designs with introduced inconsistencies ($P$-value and $O$-value threshold approach). Furthermore, for the influence diagnostics methods, we calculated the magnitude of influence for 17 designs and assessed inconsistency based on whether the design ranked among the top three in terms of influence magnitude (ranking approach). The local inconsistency test was applied to two loops involving comparisons with introduced inconsistency. Specifically, for ARB vs CT, the loops included ACE-ARB-CT (loop 1) and ARB-CCB-CT (loop 2); for CCB vs CT, the loops included ACE-CCB-CT (loop 1) and ARB-CCB-CT (loop 2). For the bootstrap methods, we performed 1,000 resamplings for individual analyses.

The results of the simulations are presented in Table 3. The local inconsistency test consistently showed inflated type I error rates ranging from 6% to 23.6%, particularly in scenarios with smaller sample sizes and higher heterogeneity. In scenarios with moderate inconsistency, the TPRs of the local inconsistency test were substantially lower than those observed in severe inconsistency scenarios (10.1%−26.5%).

For the proposed testing approach and the influence diagnostics methods using the $O$-value threshold approach, the FPRs were generally lower than those of the local inconsistency test, although they slightly exceeded the nominal 5% level, especially in scenarios with smaller sample sizes. The FPRs of the ranking approach were considerably higher than those observed for the $O$-value threshold approach. However, it is important to note that the ranking approach aims to prioritize designs for further investigation, rather than keep the FPR below a specific threshold.

In terms of the TPR, the proposed methods generally performed well and outperformed the local inconsistency test. Across all proposed methods, the TPR increased with higher inconsistency, lower heterogeneity, and larger sample sizes. Moreover, they demonstrated robust performance regardless of which design had introduced inconsistency. Among the proposed methods, MDFFITS tended to exhibit the highest TPR under both the ranking approach and the approach based on $P$-value and $O$-value thresholds. This was followed by the proposed testing approach and the averaged studentized residual. In contrast, $\Phi_d$ and $\Xi_d$ showed lower TPRs, with $\Phi_d$ performing slightly better than $\Xi_d$. Moreover, the ranking approach generally exhibited higher TPRs. While the $O$-value threshold approach showed limited performance in detecting moderate inconsistency, the ranking approach performed much better. Compared with the local inconsistency test, all proposed methods except for $\Phi_d$ and $\Xi_d$ achieved higher TPRs under the $P$-value and $O$-value threshold approach. Exceptions were scenarios 5 and 17, which involved smaller sample sizes, higher heterogeneity, and moderate inconsistency. In these scenarios, all proposed methods achieved TPRs that were more than 5% lower than that of the local inconsistency test. However, we note that the local inconsistency test exhibited FPRs exceeding 10% for scenarios 4 and 16, which were simulated under identical conditions but without inconsistency. The proposed methods, when used with the ranking approach, consistently outperformed the local inconsistency test, with TPR differences exceeding 20% in many scenarios.

## 7. Discussion

In network meta-analysis, the evaluation of inconsistency is critical for ensuring the validity of evidence synthesis. Existing methods are mainly founded on statistical testing for local or global characteristics in the network [3-7], and the results are provided, whether significant or not. The primary purpose of these analyses is to explore possible sources of bias or irregular issues in the network and prioritize the designs that have interactions with the treatments. The proposed statistical methods quantify how the overall results change when a specific set of studies is excluded, and these results will provide new insights that will enable design-by-treatment interactions to be assessed. In particular, to prioritize study designs for further investigations, these new methods will be useful alternatives to existing inconsistency evaluation tools. In addition, the simulation studies showed that, although the standard local inconsistency test could not detect inconsistencies accurately, the proposed methods detected the design-by-treatment interactions while controlling the statistical errors properly. Lu et al. [31] proposed a similar idea of using studentized residuals for assessing local inconsistency; however, they did not adopt the leave-one-design-out scheme. In general, the naïve studentized residual can be over-optimistic because it is defined by the overall estimates using the information of the target cases themselves, as is known in conventional regression diagnosis [21]. The new methods address this substantial problem and enable more sophisticated evaluations for statistical errors via a bootstrapping approach.

We also proposed a convenient summary measure: the $O$-value. The $O$-value is an effective measure for evaluating the influence of a design, accounting for the statistical variability of the influence measures. For example, even if an influence measure shows a large effect, its ranking may decrease when considering the statistical variability obtained by the $O$-value. In such cases, it indicates that, while the influence may be large, the result comes with some statistical uncertainty. The quantitative information in the influence measures is dropped in the $O$-value; however, it can be conveniently used to assess the impact of designs, similar to the $P$-value of statistical tests. The proposed influence measures and $O$-values will be effective methods for prioritizing designs. However, the results should be interpreted using both measures simultaneously. In particular, the influence measures provide relevant quantitative information concerning the degree of influence.

In addition, our methods enable direct evaluation of the influence of a design. The global inconsistency test uses whole statistical information in the network; however, it is not easy to determine where the inconsistency exists when the design-by-treatment interaction model is used. Moreover, few proposed methods can evaluate local inconsistency while addressing the influence of multi-arm studies [32,33]. Our new methods focus on assessing the influences of designs directly, and the results indicate which designs influence consistency in the network. Furthermore, the global inconsistency test rarely provides significant results, whereas our methods detect potentially relevant local information effectively, as demonstrated in Section 5. These results will provide alternative new insights that were not provided by existing methods.

As shown in the simulation studies, the local inconsistency test could not detect inconsistency in a network in practical situations. It showed inflated type I error rates and low TPRs under moderate inconsistency conditions. The proposed methods generally outperformed the local inconsistency test regardless of sample size, degree of heterogeneity, or the presence of multi-arm studies. Although the performances of the proposed methods tended to decline in situations with smaller sample sizes, high heterogeneity, and moderate inconsistency (particularly $\Phi_d$ and $\Xi_d$), the ranking approach offered a higher likelihood of identifying influential designs. Although the FPR of the ranking approach was higher, TPR should be emphasized over FPR, given that the primary goal is to prioritize designs for further investigation. In practice, the prioritization of designs depends on the context; however, these statistical outputs will provide useful information for further investigation.

Among the four influence diagnostics methods, MDFFITS demonstrated the highest TPR, though it had a slightly higher FPR. However, because the proposed measures reflect different aspects of influence of a design and collectively offer a comprehensive picture, we recommend using all four diagnostic methods to evaluate inconsistency, incorporating both influence magnitudes and $O$-values, regardless of the sample size, degree of heterogeneity, or presence of multi-arm studies. Additionally, when adopting a hypothesis testing framework to assess inconsistency, we suggest using the proposed testing approach, which generally exhibited lower type I error rates and higher TPRs than the local inconsistency test.

One potential limitation of the proposed methods is that their performance may be suboptimal when multiple inconsistent designs are present. In such cases, the leave-one-out dataset may still contain inconsistent designs. This is a common drawback of the leave-one-out approach, known as the masking effect or swamping effect [18,23]. Several solutions have been proposed, and a sequential algorithm, for example, could be an effective approach [34,35]. Extending the proposed methods for incorporation into these advanced frameworks is an important subject for future research. Another potential limitation is the computational cost associated with the bootstrapping method. When the number of designs is large, the computational burden can be large. However, with modern computational environments, this is unlikely to be a major practical issue.

In conclusion, our proposed methods are expected to provide new insights in the inconsistency evaluation of network meta-analysis. In addition to the current standard testing-based approaches, these methods can be used as alternative effective approaches for locating inconsistency in a network.

## Acknowledgements

This study was supported by Grants-in-Aid for Scientific Research from the Japan Society for the Promotion of Science (Grant numbers: JP23K24811, JP24K21306, JP22K19688, and JP23K11931).

**Table 1**. Summary of the network meta-analysis dataset of Sciarretta et al. [12]

| Design ID | Design | Trial | Drug 1 | d/n | Drug 2 | d/n | Drug 3 | d/n |
|---|---|---|---|---|---|---|---|---|
| 1 | ACE vs ARB | ONTARGET | ACE | 514/8576 | ARB | 537/8542 | — | — |
| 2 | ACE vs BB | UKPDS | ACE | 12/400 | BB | 9/358 | — | — |
| 3 | ACE vs CCB | ABCD | ACE | 5/235 | CCB | 6/235 | — | — |
| 4 | ACE vs CCB vs CT | STOP-2 | ACE | 149/2205 | CCB | 186/2196 | CT | 177/2213 |
| 5 | ACE vs CCB vs DD | ALLHAT2002 | ACE | 612/9054 | CCB | 706/9048 | DD | 870/15255 |
| 6 | ACE vs CT | CAPPP | ACE | 75/5492 | CT | 66/5493 | — | — |
| 7 | ACE vs DD | ANBP2 | ACE | 69/3044 | DD | 78/3039 | — | — |
| 8 | ACE vs Placebo | HOPE | ACE | 417/4645 | Placebo | 535/4652 | — | — |
| 9 | ARB vs BB | LIFE | ARB | 153/4605 | BB | 161/4588 | — | — |
| 10 | ARB vs CCB | VALUE | ARB | 354/7649 | CCB | 400/7596 | — | — |
| 11 | ARB vs CT | E-COST | ARB | 35/1053 | CT | 41/995 | — | — |
| 11 | ARB vs CT | Jikei Heart Study | ARB | 19/1541 | CT | 36/1540 | — | — |
| 12 | ARB vs Placebo | RENRAL | ARB | 89/751 | Placebo | 127/762 | — | — |
| 12 | ARB vs Placebo | TRANSEND | ARB | 134/2954 | Placebo | 129/2972 | — | — |
| 13 | BB vs CCB | ASCOT-BPLA | BB | 159/9618 | CCB | 134/9639 | — | — |
| 14 | CCB vs CT | NORDIL | CCB | 63/5410 | CT | 53/5471 | — | — |
| 14 | CCB vs CT | CONVINCE | CCB | 126/8179 | CT | 100/8297 | — | — |
| 15 | CCB vs DD | VHAS | CCB | 2/707 | DD | 0/707 | — | — |
| 15 | CCB vs DD | NICS-EH | CCB | 0/204 | DD | 3/210 | — | — |
| 15 | CCB vs DD | INSIGHT | CCB | 26/3157 | DD | 12/3164 | — | — |
| 15 | CCB vs DD | SHELL | CCB | 23/942 | DD | 19/940 | — | — |
| 16 | CCB vs Placebo | Syst-Eur | CCB | 37/2398 | Placebo | 49/2297 | — | — |
| 16 | CCB vs Placebo | Syst-China | CCB | 4/1253 | Placebo | 8/1141 | — | — |
| 16 | CCB vs Placebo | FEVER | CCB | 18/4841 | Placebo | 27/4870 | — | — |
| 17 | DD vs Placebo | HYVET | DD | 22/1933 | Placebo | 57/1912 | — | — |
| 18 | AB vs DD | ALLHAT2000 | AB | 491/9067 | DD | 420/15268 | — | — |

Abbreviations: AB, α-blocker; ACE, angiotensin-converting enzyme inhibitor; ARB, angiotensin II receptor blocker; BB, β-blocker; CCB, calcium channel blocker; CT, conventional treatment; DD, diuretic; d/n, number of incidences of heart failure/number of participants.

**Table 2.** Antihypertensive drugs network meta-analysis results of the proposed testing approach.

| Design ID | Design | $W_d$ | $P$-value | Treatment | Reference treatment | $\exp(\hat{\mu}^{(T_d)})$ * | $\exp(\hat{\mu}^{(-T_d)})$ * |
|---|---|---|---|---|---|---|---|
| 11 | ARB vs CT | 4.472 | 0.039 | CT | ARB | 1.501 | 0.911 |
| 17 | DD vs Placebo | 4.216 | 0.056 | Placebo | DD | 2.669 | 1.534 |
| 7 | ACE vs DD | 3.743 | 0.091 | DD | ACE | 1.136 | 0.787 |
| 16 | CCB vs Placebo | 2.125 | 0.141 | Placebo | CCB | 1.474 | 1.107 |
| 14 | CCB vs CT | 2.306 | 0.141 | CT | CCB | 0.797 | 1.010 |
| 13 | BB vs CCB | 2.093 | 0.212 | CCB | BB | 0.839 | 1.115 |
| 6 | ACE vs CT | 1.803 | 0.227 | CT | ACE | 0.878 | 1.144 |
| 2 | ACE vs BB | 0.805 | 0.384 | BB | ACE | 0.834 | 1.265 |
| 9 | ARB vs BB | 0.951 | 0.394 | BB | ARB | 1.058 | 1.286 |
| 12 | ARB vs Placebo | 0.704 | 0.419 | Placebo | ARB | 1.183 | 1.455 |
| 8 | ACE vs Placebo | 0.835 | 0.467 | Placebo | ACE | 1.318 | 1.502 |
| 15 | CCB vs DD | 0.099 | 0.749 | DD | CCB | 0.656 | 0.721 |
| 4 | ACE vs CCB vs CT | 0.474 | 0.786 | CCB | ACE | 1.277 | 1.152 |
| | | | | CT | ACE | 1.200 | 1.044 |
| 10 | ARB vs CCB | 0.110 | 0.803 | CCB | ARB | 1.145 | 1.097 |
| 1 | ACE vs ARB | 0.006 | 0.955 | ARB | ACE | 1.052 | 1.063 |
| 3 | ACE vs CCB | 0.001 | 0.975 | CCB | ACE | 1.205 | 1.181 |
| 5 | ACE vs CCB vs DD | 0.007 | 0.996 | CCB | ACE | 1.167 | 1.183 |
| | | | | DD | ACE | 0.834 | 0.831 |
| 18 | AB vs DD | — | — | — | — | — | — |

* $\hat{\mu}$ was estimated using the log odds ratio (OR) scale, and the estimate was transformed into the OR scale.

Abbreviations: AB, α-blocker; ACE, angiotensin-converting enzyme inhibitor; ARB, angiotensin II receptor blocker; BB, β-blocker; CCB, calcium channel blocker; CT, conventional treatment; DD, diuretic.

**Table 3.** Simulation results: True positive rates (TPRs) and false positive rates (FPRs) [%] for inconsistency evaluation.

| Scenario | Design with introduced inconsistency | $N$ | $\tau$ | $\omega^*$ | Proposed methods | | | | | | | | | Inconsistency test | |
|---|---|---|---|---|---|---|---|---|---|---|---|---|---|---|---|
| | | | | | Averaged studentized residual ($\Psi_d$) | | MDFFITS | | $\Phi_d$ | | $\Xi_d$ | | Testing approach | Local inconsistency test | |
| | | | | | $O$-value $< 0.05$[†] | Top 3[‡] | $O$-value $< 0.05$[†] | Top 3[‡] | $O$-value $< 0.05$[†] | Top 3[‡] | $O$-value $< 0.05$[†] | Top 3[‡] | $P$-value $< 0.05$[†] | $P$-value $< 0.05$[†] Loop 1[§] | $P$-value $< 0.05$[†] Loop 2[§] |
| 1 | ARB vs CT | 26 | 0.05 | 0 | 5.7 | 19.0 | 6.3 | 24.8 | 5.6 | 20.1 | 5.5 | 21.2 | 7.3 | 8.4 | 6.4 |
| 2 | | | | −0.1 | 16.4 | 35.6 | 18.7 | 43.6 | 12.4 | 34.0 | 10.8 | 35.1 | 17.6 | 13.0 | 15.3 |
| 3 | | | | −0.3 | 79.6 | 89.2 | 83.2 | 92.6 | 68.4 | 88.5 | 61.8 | 88.1 | 77.7 | 51.1 | 62.4 |
| 4 | | | 0.1 | 0 | 4.1 | 16.4 | 4.5 | 24.4 | 3.5 | 20.4 | 2.8 | 21.1 | 5.5 | 12.1 | 11.6 |
| 5 | | | | −0.1 | 10.4 | 27.2 | 11.5 | 37.7 | 7.2 | 30.9 | 6.0 | 29.6 | 12.3 | 16.7 | 17.9 |
| 6 | | | | −0.3 | 50.5 | 72.5 | 56.4 | 80.8 | 39.5 | 71.9 | 34.0 | 68.9 | 51.0 | 40.6 | 41.7 |
| 7 | | 52 | 0.05 | 0 | 5.5 | 16.2 | 6.0 | 26.4 | 4.4 | 22.3 | 3.9 | 22.4 | 6.3 | 7.5 | 6.0 |
| 8 | | | | −0.1 | 27.9 | 46.1 | 35.0 | 63.6 | 17.7 | 45.8 | 15.6 | 44.9 | 35.6 | 22.6 | 25.4 |
| 9 | | | | −0.3 | 97.9 | 98.9 | 98.7 | 99.3 | 94.2 | 98.3 | 93.1 | 98.0 | 96.8 | 84.2 | 89.1 |
| 10 | | | 0.1 | 0 | 3.9 | 16.5 | 4.6 | 23.1 | 4.3 | 20.9 | 3.2 | 20.5 | 5.2 | 9.3 | 9.7 |
| 11 | | | | −0.1 | 14.8 | 30.0 | 18.7 | 47.9 | 9.4 | 33.5 | 8.5 | 31.4 | 18.1 | 15.9 | 19.1 |
| 12 | | | | −0.3 | 80.2 | 90.2 | 88.9 | 97.3 | 68.7 | 88.2 | 65.6 | 84.6 | 84.4 | 62.4 | 67.3 |
| 13 | CCB vs CT | 26 | 0.05 | 0 | 5.9 | 19.0 | 6.6 | 21.5 | 5.1 | 20.1 | 5.1 | 21.7 | 6.4 | 11.7 | 6.8 |
| 14 | | | | −0.1 | 21.0 | 38.0 | 22.5 | 44.8 | 15.4 | 38.2 | 14.2 | 39.4 | 21.6 | 20.8 | 10.1 |
| 15 | | | | −0.3 | 86.4 | 93.0 | 88.5 | 93.1 | 76.6 | 91.4 | 72.3 | 91.0 | 79.5 | 65.8 | 12.3 |
| 16 | | | 0.1 | 0 | 6.2 | 18.7 | 6.5 | 23.4 | 4.9 | 22.0 | 3.8 | 22.0 | 6.6 | 23.6 | 11.0 |
| 17 | | | | −0.1 | 11.0 | 27.4 | 11.2 | 33.1 | 8.7 | 29.8 | 7.1 | 28.2 | 11.1 | 24.9 | 13.4 |
| 18 | | | | −0.3 | 57.6 | 77.2 | 62.1 | 82.7 | 48.2 | 79.1 | 42.0 | 76.3 | 57.0 | 51.8 | 14.0 |
| 19 | | 52 | 0.05 | 0 | 4.6 | 14.8 | 5.1 | 19.6 | 3.1 | 19.2 | 2.8 | 20.1 | 4.3 | 10.3 | 7.1 |
| 20 | | | | −0.1 | 28.7 | 49.1 | 39.8 | 63.0 | 20.0 | 50.1 | 18.1 | 49.9 | 36.5 | 26.5 | 11.6 |
| 21 | | | | −0.3 | 97.7 | 98.8 | 99.3 | 99.3 | 96.0 | 98.7 | 95.3 | 98.5 | 97.9 | 92.0 | 23.9 |
| 22 | | | 0.1 | 0 | 4.5 | 17.1 | 4.6 | 20.7 | 3.8 | 20.4 | 3.3 | 20.7 | 4.8 | 14.0 | 8.3 |
| 23 | | | | −0.1 | 15.3 | 32.6 | 20.2 | 44.1 | 10.9 | 34.9 | 9.8 | 33.9 | 18.5 | 21.2 | 12.8 |
| 24 | | | | −0.3 | 85.3 | 93.8 | 92.4 | 97.2 | 76.7 | 92.4 | 74.4 | 89.8 | 85.3 | 74.1 | 20.6 |

[*] $\omega$ is the inconsistency factor, where 0 indicates no inconsistency, −0.1 indicates moderate inconsistency, and −0.3 indicates severe inconsistency.

[†] The results are presented as the rate (%) of $P$-value ($O$-value) results $< 0.05$ for the designs, representing the TPR for scenarios with inconsistency ($\omega = -0.1$ or $-0.3$) and FPR for scenarios with no inconsistency ($\omega = 0$).

[‡] The results are presented as the rate (%) at which the design was ranked among the top-three most influential according to each influence diagnostic measure, representing the TPR when $\omega$ is $-0.1$ or $-0.3$, and the FPR when $\omega$ is 0.

[§] Loop 1 refers to the ACE-ARB-CT loop for scenarios 1−12 and the ACE-CCB-CT loop for scenarios 13−24. Loop 2 refers to the ARB-CCB-CT loop for scenarios 1−24.

Note: TPRs are presented in bold font for clarity.

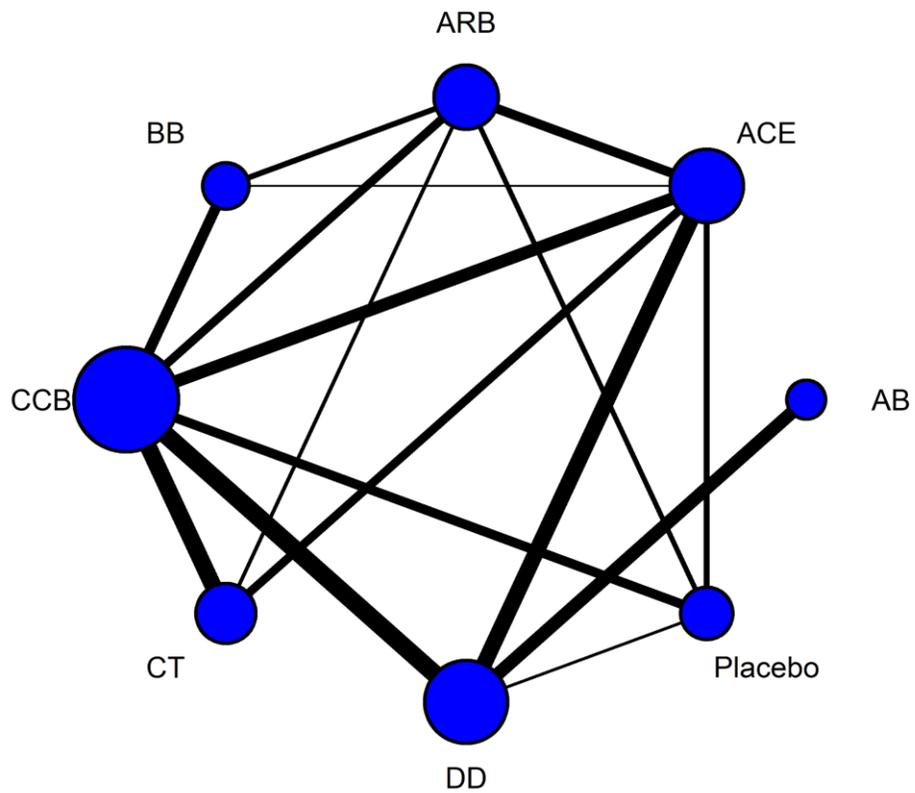

**Figure 1.** Network diagram of the network meta-analysis for antihypertensive drugs. (AB, α-blocker; ACE, angiotensin-converting enzyme inhibitor; ARB, angiotensin II receptor blocker; BB, β-blocker; CCB, calcium channel blocker; CT, conventional treatment; DD, diuretic)

**(a)**

| Treatment | | OR | 95%CI |
|---|---|---|---|
| DD | | 0.600 | (0.487, 0.739) |
| ACE | | 0.714 | (0.607, 0.840) |
| ARB | | 0.759 | (0.642, 0.896) |
| CT | | 0.775 | (0.626, 0.959) |
| CCB | | 0.843 | (0.709, 1.004) |
| BB | | 0.878 | (0.679, 1.134) |
| AB | | 1.214 | (0.887, 1.662) |

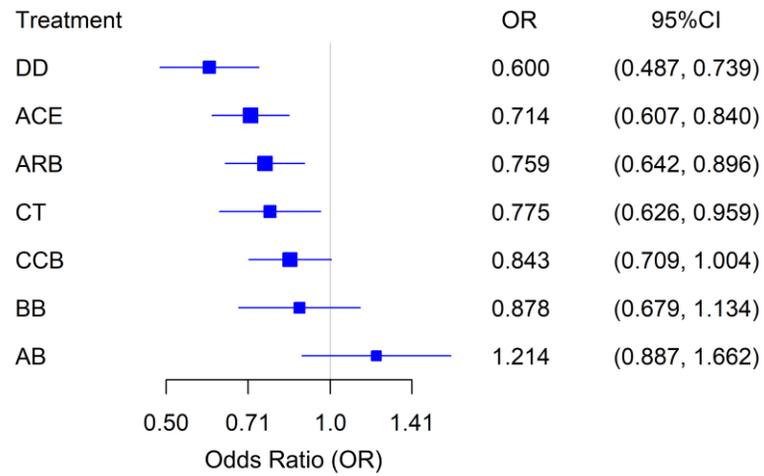

**(b)**

| Treatment | | OR | 95%CI |
|---|---|---|---|
| DD | | 0.643 | (0.541, 0.765) |
| ACE | | 0.745 | (0.653, 0.850) |
| CT | | 0.759 | (0.625, 0.921) |
| ARB | | 0.811 | (0.704, 0.933) |
| CCB | | 0.879 | (0.759, 1.019) |
| BB | | 0.931 | (0.749, 1.157) |
| AB | | 1.302 | (1.021, 1.660) |

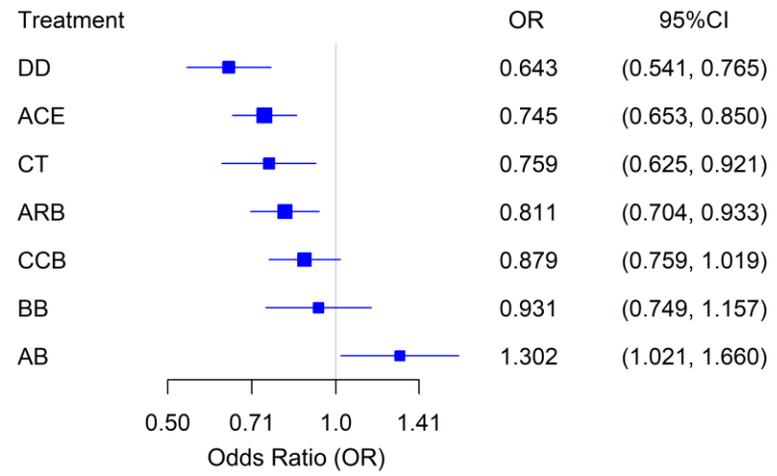

**Figure 2.** Pooled odds ratio (OR) estimates of antihypertensive drugs compared with a placebo: (a) for all 26 studies; (b) for 23 studies after excluding the Jikei Heart Study, E-COST, and HYVET. (AB, α-blocker; ACE, angiotensin-converting enzyme inhibitor; ARB, angiotensin II receptor blocker; BB, β-blocker; CCB, calcium channel blocker; CT, conventional treatment; DD, diuretic)

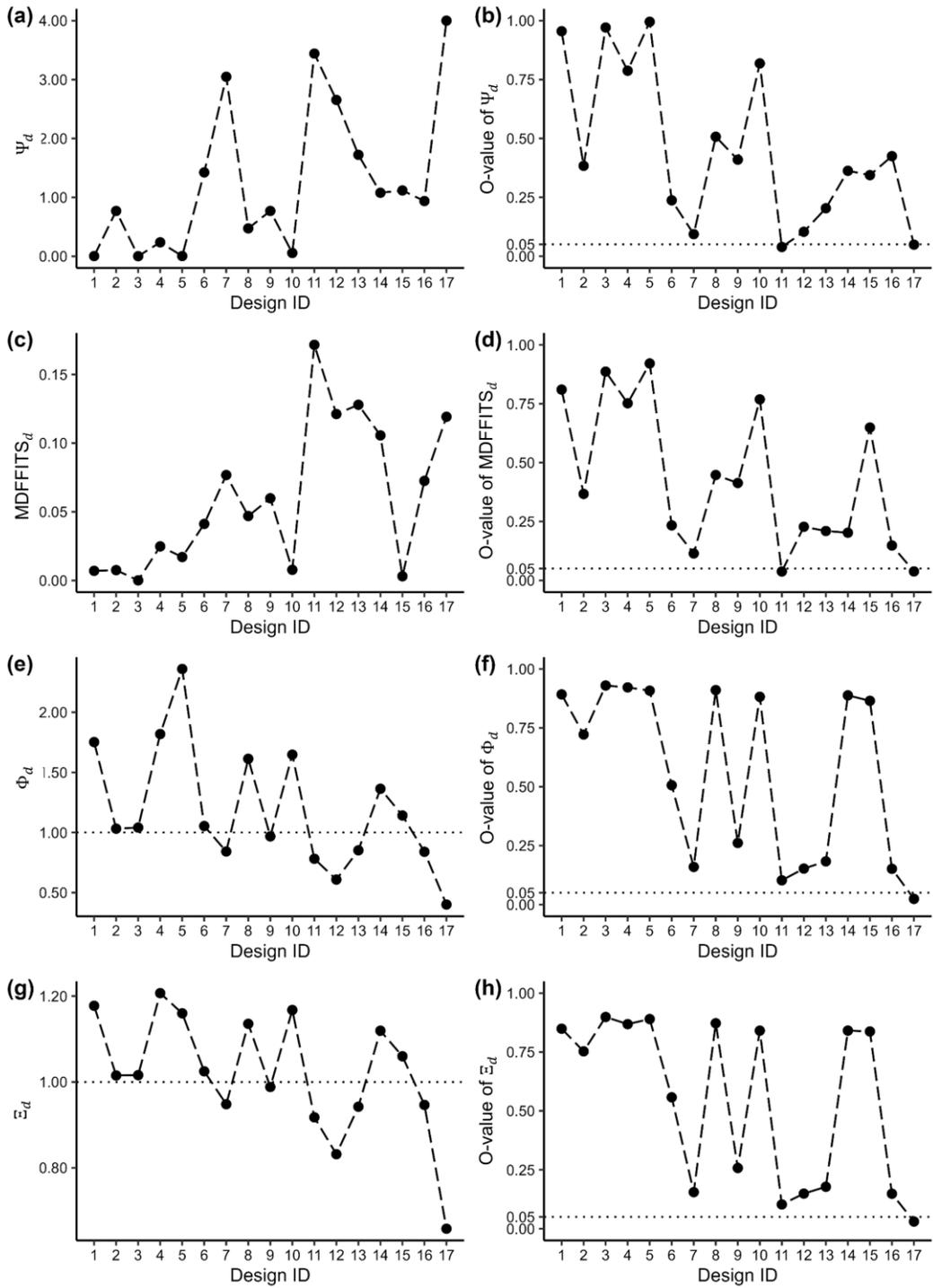

**Figure 3.** Influence diagnostics results for evaluating inconsistency: (a) $\Psi_d$ (averaged studentized residual); (b) $O$-value of $\Psi_d$; (c) $\mathrm{MDFFITS}_d$; (d) $O$-value of $\mathrm{MDFFITS}_d$; (e) $\Phi_d$; (f) $O$-value of $\Phi_d$; (g) $\Xi_d$; (h) $O$-value of $\Xi_d$.